\documentclass[aps,prd,english,superscriptaddress,11pt,notitlepage]{revtex4}
	\usepackage[colorlinks=true, a4paper=true, pdfstartview=FitV,
linkcolor=blue, citecolor=blue, urlcolor=blue]{hyperref}

\usepackage{amsmath}
\usepackage{amssymb}
\usepackage{graphicx}
\usepackage{hhline}
\usepackage{textcomp}
\makeatletter
\usepackage{babel}
\newcommand{\bea}{\begin{eqnarray}}
\newcommand{\eea}{\end{eqnarray}}

\begin{document}


\title{Collective coordinate model of kink-antikink collisions in $\phi^4$ theory}

\author{N.~S. Manton}

\affiliation{Department of Applied Mathematics and Theoretical Physics,
University of Cambridge,
Wilberforce Road, Cambridge CB3 0WA, U.K.}

\author{K. Ole\'{s}}%

\author{T. Roma\'{n}czukiewicz}

\author{A. Wereszczy\'{n}ski}

\affiliation{
 Institute of Theoretical Physics,  Jagiellonian University, Lojasiewicza 11, 30-348 Krak\'{o}w, Poland
}

\date{\today}

\begin{abstract}
The fractal velocity pattern in symmetric kink-antikink collisions in $\phi^4$ 
theory is shown to emerge from a dynamical model with two effective moduli, the
kink-antikink separation and the internal shape mode amplitude. The
shape mode usefully approximates Lorentz contractions of the kink and antikink, 
and the previously problematic null-vector in the shape mode amplitude at 
zero separation is regularized.
\end{abstract}

\pacs{Valid PACS appear here}
\maketitle


\section{\label{sec:motiv}Introduction}

Despite the frequent occurrence of topological solitons in Nature, and
their theoretical importance, the collision and scattering of solitons 
in non-integrable field theories is far from fully understood. Even in 
the prototypical case of kink-antikink (KAK) collisions in
$\phi^4$ theory in 1+1 dimensions, there is little understanding
of the intriguing fractal pattern of final velocities, alternating with
regions of KAK annihilation, as the initial velocities vary 
\cite{Sug,Mosh,CSW}. Although the role of resonant energy
transfer between the translational and vibrational 
degrees of freedom (DoF) of the solitons has been emphasized, no 
detailed effective model with finitely many DoF has been derived from 
the field theory Lagrangian, despite four decades of 
investigation \cite{KG}. Sugiyama's original attempt \cite{Sug},
studied by many others \cite{Per}, appears after the correction of a 
typographical error by Takyi and Weigel \cite{TW} to lead to wrong 
predictions. However, building on our recent work \cite{MORW}, we will show 
that the problems are technical and can be overcome. 

An effective model truncates the $\phi^4$ field theory Lagrangian
\begin{equation}
L[\phi]=\int_{-\infty}^{+\infty} \left( \frac{1}{2} \dot{\phi}^2 -
  \frac{1}{2} \phi'^2 - \frac{1}{2}(1-\phi^2)^2 \right) dx
\label{Lagfield}
\end{equation}
to a Lagrangian dynamics of collective coordinates or {\it moduli} $X = 
\{X^i, i=1,\dots,N \}$. Field configuration space is judiciously
reduced to a finite-dimensional subspace, the {\it moduli space} 
$\mathcal{M}[X]=\{ \widetilde{\phi}(x;X) \}$, which represents, for example, 
KAK superpositions with the separation as modulus. 

Implementation requires the configurations $\widetilde{\phi}(x;X(t))$ 
to be inserted into (\ref{Lagfield}), and the integral over $x$ 
performed. The result is a (non-relativistic) effective Lagrangian
on moduli space
\begin{equation}
L=T-V=\frac{1}{2} g_{ij}(X) \dot{X}^i \dot{X}^j - V(X) \,,
\label{EffL} 
\end{equation}
where the metric $g_{ij}$ inherited from the kinetic terms in
(\ref{Lagfield}) is 
\begin{equation}
g_{ij}(X)=\int_{-\infty}^{+\infty}  \frac{\partial \widetilde{\phi}
  (x;X)}{\partial X^i} \frac{\partial \widetilde{\phi} (x;X)}{\partial
  X^j} \, dx 
\label{EffMet}
\end{equation}
and generally curved, and the potential $V$ is
\begin{equation}
V(X)=\int_{-\infty}^{+\infty} \left( \frac{1}{2} \tilde\phi'(x;X)^2 
+ \frac{1}{2}\bigl(1-\widetilde\phi(x;X)^2\bigr)^2 \right) dx \,.
\label{EffPot}
\end{equation}
The field equations are then approximated by the Euler--Lagrange 
equations derived from (\ref{EffL}),
\begin{equation}
g_{ij}(X) \bigl( \ddot{X}^j + \Gamma^j_{kl}(X) \dot{X}^k \dot{X}^l \bigr) +
\frac{\partial V}{\partial X^i} = 0 \,,
\end{equation}
where $\Gamma^j_{kl}$ is the Levi-Civita connection of the
metric. This is a system of ODEs. 

In contrast to the Bogomol'nyi--Prasad--Sommerfield (BPS)
situation, where the reduced dynamics is accurately described by
geodesic flow on the {\it canonical moduli space} of
minimal-energy soliton solutions \cite{M,AH,S}, there
is no unique moduli space of KAK configurations. However, it is 
agreed that the collective coordinates should be related to the 
lowest-frequency excitations of static kinks.
 
In $\phi^4$ theory, there are two such excitations solving the 
linearized field equation around the kink $\phi_{\rm K}(x) = \tanh(x)$ -- 
the zero (frequency) mode $\phi'_{\rm K}(x) = 1/\cosh^2(x)$
arising from translational invariance, and the normalizable shape mode
\begin{equation}
\eta(x)= \frac{\sinh(x)}{\cosh^2(x)}
\end{equation}
with frequency $\omega = \sqrt{3}$, just below 
the continuum starting at $\omega = 2$. Therefore, useful single-kink 
configurations have a kink at location $a$, excited by its shape
mode with amplitude $A$. The moduli space dynamics of $a$ and $A$ 
models kink dynamics well, but not exactly because $A$ is finite
rather than infinitesimal, and nonlinear effects including radiation
from the vibrating shape mode are neglected. The single-kink sector
provides the initial data for KAK collisions. 

The antikink $\phi_{\rm AK}(x) = -\tanh(x)$ has the same two modes, and
KAK dynamics can be modelled by superposing kink and
antikink fields, as described below. The effective model for 
reflection-symmetric KAK collisions is a non-integrable Lagrangian 
system with two DoF, like a double pendulum, and is found 
to agree in many details with the full field theory dynamics. 
This resolves the long-standing problem connected with the KAK 
system, and confirms that resonant energy transfer between 
the relative translational motion and shape vibrations is responsible
for the observed fractal structure. From a wider perspective, we see 
that collective coordinate dynamics can be a very useful tool for 
non-integrable solitons.
 
\section{\label{sec:kink}Vibrating Kink}

There is a canonical moduli space of static kinks 
$\widetilde{\phi}(x;a)= \tanh(x-a)$ having energy (mass) $4/3$ and 
solving the Bogomolny equation $\phi' = 1 - \phi^2$, with the kink 
location $a$ as modulus. This extends to a 2-dimensional moduli 
space of kinks deformed by the shape mode,
\begin{equation}
\widetilde{\phi}(x;a,A) = \tanh(x-a) + A \frac{\sinh(x-a)}{\cosh^2(x-a)} \,.
\end{equation}
Treating $a$ and $A$ as time-dependent and substituting into 
(\ref{Lagfield}) gives an effective Lagrangian for a moving, 
vibrating kink of the form
\begin{equation}
L_{\rm K}[a,A]=\frac{1}{2}g_{aa}(A)\dot{a}^2+\frac{1}{2}g_{AA}\dot{A}^2-V(A) \,.
\end{equation}
The kinetic terms define a diagonal, wormhole metric on the moduli 
space, with components
\begin{equation}
g_{aa}(A)=\frac{4}{3} +\frac{\pi}{2} A+\frac{14}{15}A^2 \,, \quad
g_{AA}=\frac{2}{3} \,,
\end{equation}
and the potential is
\begin{equation}
V(A)=\frac{4}{3} + A^2+\frac{\pi}{8} A^3 +\frac{2}{35}A^4 \,.
\end{equation}
A 2-dimensional wormhole is a pair of planes smoothly connected by a 
curved throat; here, the throat is located at $A \approx -0.84$, where
$g_{aa}$ is minimal and the curvature is maximal. $a$ would normally be an 
angular variable, but $a$ has infinite range here, so the moduli space 
is the universal cover of the wormhole. Note that $V(A)$ is not 
symmetric with respect to the throat location.

The vibrationally excited kink motion is modelled by the ODE dynamics
\begin{eqnarray}
\frac{d}{dt} \left[ \left( \frac{4}{3} +\frac{\pi}{2} A  +
  \frac{14}{15} A^2 \right)  \dot{a} \right]&=&0 \,, \\
\frac{2}{3} \ddot{A} - \frac{1}{2}\left( \frac{\pi}{2}  +
  2\frac{14}{15} A \right)  \dot{a}^2 + \frac{dV}{dA}&=&0 \,, \label{Addot} 
\end{eqnarray}
which can be integrated using the conserved momentum and energy
\begin{eqnarray}
P&=&\left(\frac{4}{3} +\frac{\pi}{2} A  + \frac{14}{15} A^2 \right)
     \dot{a} \,, \\
E&=& \frac{1}{3} \dot{A}^2+ \frac{P^2}{2\left( \frac{4}{3}
     +\frac{\pi}{2} A  + \frac{14}{15} A^2\right) } + V(A) \,.
\end{eqnarray}

A key observation is that there is a stationary solution where the 
kink moves with constant velocity $\dot{a}=v$, and a constant 
shape mode amplitude
\begin{equation}
A = \frac{\pi}{8}v^2
+ \frac{\pi}{8}\left(\frac{7}{15} - \frac{3\pi^2}{128} \right)v^4
+ O(v^6)
\label{v-A}
\end{equation}
obtained by solving (\ref{Addot}) with $\ddot{A} = 0$.
The non-zero amplitude represents an approximate Lorentz contraction 
of the kink. Indeed, the exact moving kink solution 
$\phi(x,t) = \tanh\bigl((x-vt)/\sqrt{1-v^2}\bigr)$ has the 
expansion for small $v$
\begin{equation}
\phi(x,t) =  \tanh (x-vt) + \frac{v^2}{2}\frac{x-vt}{\cosh^2(x-vt)} \,.
\end{equation}
The function in the second term is the {\it Derrick mode}, 
arising from infinitesimal rescaling of the kink. The
normalized shape mode and Derrick mode, respectively
\begin{equation}
\eta(x) = \sqrt{\frac{3}{2}} \frac{\sinh(x)}{\cosh^2(x)} \quad 
{\rm and} \quad \eta_D(x)= \frac{3}{\sqrt{\pi^2-6}}
\frac{x}{\cosh^2(x)} \,,
\end{equation}
have inner product
$(\eta, \eta_D)=\pi \sqrt{3/(8(\pi^2-6))}\approx 0.98$,
so they are very similar. At $O(v^2)$, the coefficients of
these normalized modes for the stationary, moving kink have the ratio
\begin{equation}
\left(\frac{\pi v^2}{8}\sqrt{\frac{2}{3}}\right) \, \bigg/ \,
\left(\frac{v^2}{2}\frac{\sqrt{\pi^2-6}}{3}\right) \approx 0.98 \,,
\end{equation}
similarly close to 1. So the shape mode Lorentz contracts the 
kink to a good approximation. This can be exploited in initial 
KAK collision data. Further oscillations of the shape mode 
describe a vibrating kink in motion, approximating a Lorentz-boosted,
vibrating kink at rest.

\section{\label{sec:moduli}Kink-Antikink Moduli Space}

\begin{figure}
\includegraphics[width=0.85\columnwidth]{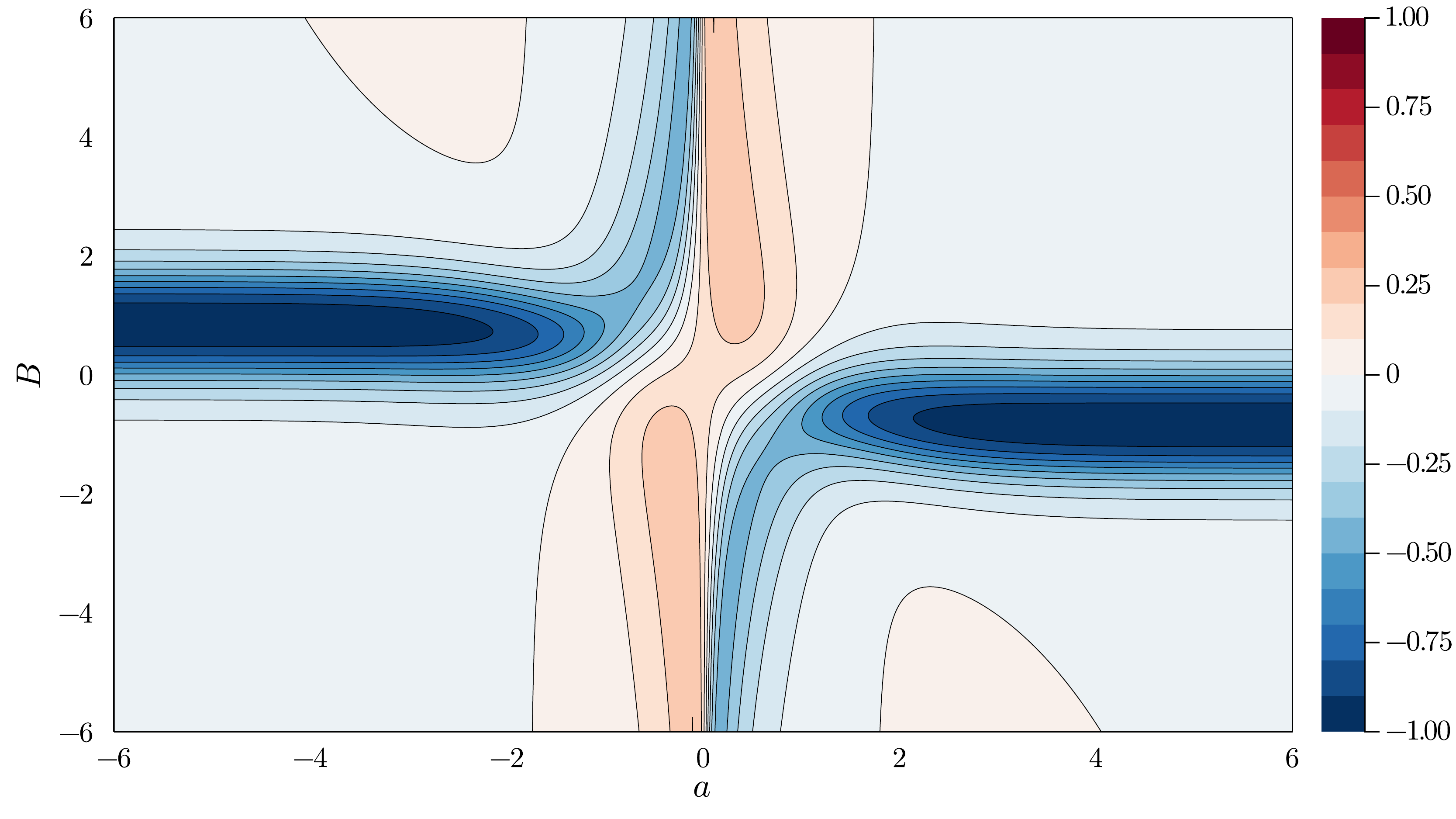}
 \caption{Tanh of the Ricci scalar curvature}\label{fig:curv}
 \vspace*{0.2cm}
\hspace*{-0.5cm} \includegraphics[width=0.85\columnwidth]{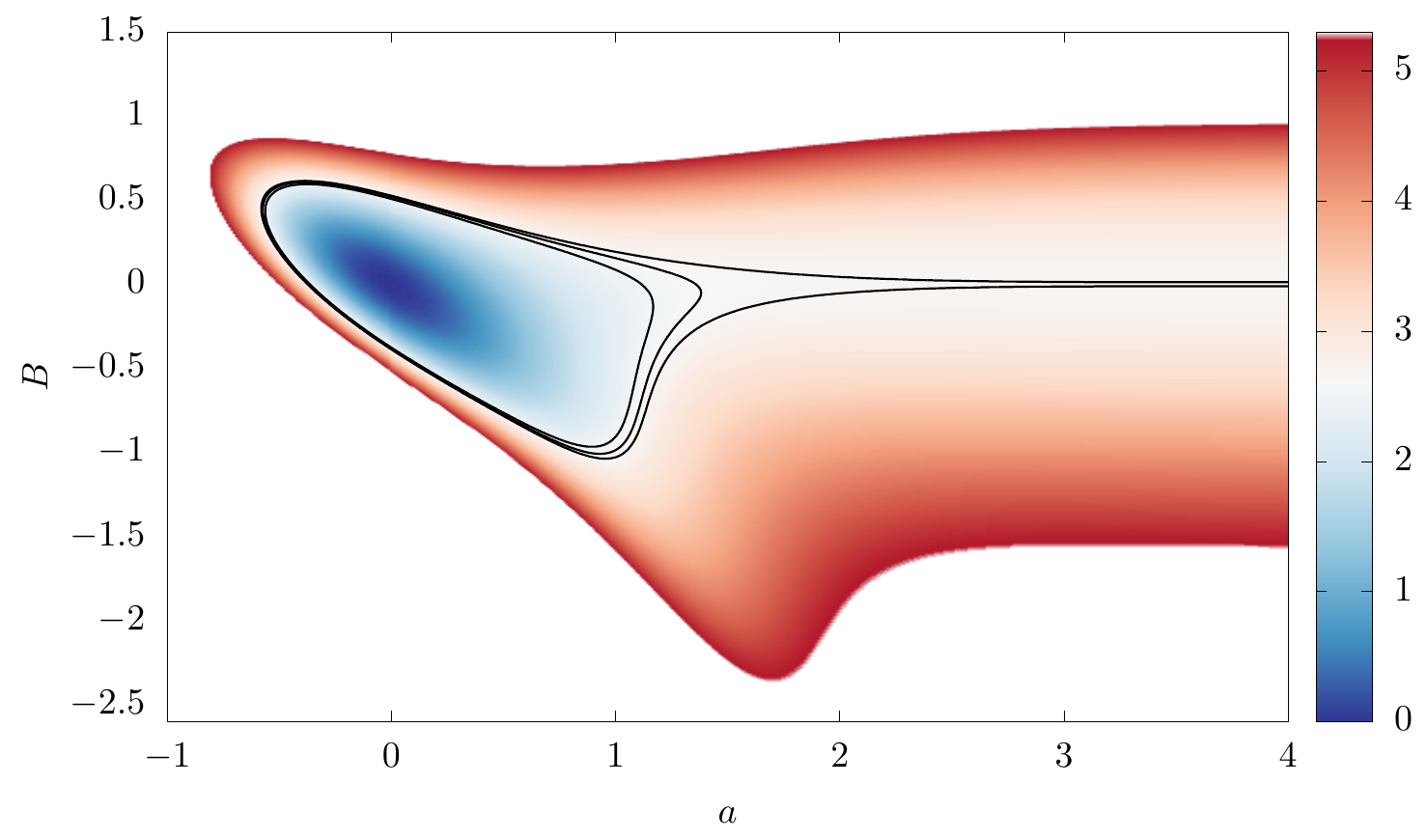}
\caption{The potential $V(a,B)$ with equipotential contours:
$V=2.5, 2.6, 2.667$ }\label{fig:pot}
\end{figure}
Following Sugiyama \cite{Sug}, we model KAK configurations as 
simple kink-antikink superpositions, with a constant field 
shift by 1 to satisfy the boundary conditions. We focus on 
configurations with reflection symmetry about $x=0$, so assume 
the kink and antikink are located at $\mp a$ and have equal
shape mode amplitudes $A$. These configurations are 
\begin{equation}
\widetilde{\phi}(x; a,A)=\tanh(x+a) - \tanh(x-a) -1 
+ A \left( \frac{\sinh(x+a)}{\cosh^2(x+a)} 
- \frac{\sinh(x-a)}{\cosh^2(x-a)} \right) \,, 
\label{KAKnaive}
\end{equation}
and they define a 2-dimensional moduli space. The shape modes
are not deformed by the presence of the partner, so we say 
they are frozen, even though their amplitudes are not. 
The main disadvantage of the formula (\ref{KAKnaive}) is the null-vector
problem \cite{TW}. Because $\partial \widetilde{\phi} / \partial A
|_{a=0}=0$, the components $g_{AA}$ and $g_{aA}$ of the moduli 
space metric vanish at $a=0$, so $A$ is not globally a good
coordinate. This leads to apparent singularities in the
moduli space dynamics.
 
This problem is resolved by redefining the coordinate $A$ as 
$B/\tanh(a)$ \cite{MORW}. Then
\begin{equation}
\widetilde{\phi}(x; a,B)=\tanh(x+a) - \tanh(x-a) -1 
+  \frac{B}{\tanh(a)} 
\left( \frac{\sinh(x+a)}{\cosh^2(x+a)} 
- \frac{\sinh(x-a)}{\cosh^2(x-a)} \right)
\label{KAKrefined}
\end{equation}
is a moduli space of field configurations with coordinates 
$(a,B) \in \mathbb{R}^2$ having a globally well defined metric and
potential. In particular, for $a$ close to 0,
\begin{equation}
\widetilde{\phi}(x;a,B) \approx \frac{2a}{\cosh^{2}(x)}+2B\left(\frac{2}{\cosh^{3}(x)} -
\frac{1}{\cosh(x)}\right)-1 \,,
\end{equation} 
a linear expression in both coordinates. The 
derivatives of $\widetilde{\phi}$ w.r.t. $a$ and $B$ are non-vanishing 
functions, which solves the null-vector problem.
 
The effective model has the form (\ref{EffL}) with $(X^1, X^2)
= (a,B)$. The non-diagonal metric $g_{ij}$ can be determined analytically 
from the integrals (\ref{EffMet}), and the potential $V$ from (\ref{EffPot}), 
\bea
g_{aa}&=& \frac{2}{3\sinh^3(2 a)}
\bigl(-15 \sinh (2 a)+\sinh (6 a)+24 a \cosh (2 a)\bigr)
+ \pi B \tanh ^3(a) \nonumber \\
&+& B^2 \left( \frac{28}{15} + \frac{14}{\sinh^6(a)}\bigl(-1+a \coth (a)\bigr)
+ \frac{1}{\sinh^4(a)} \left(-\frac{50}{3}+ 12 a \coth (a)\right)
\right. \nonumber \\
&+& \left. \frac{1}{\sinh^2(a)}\left(-\frac{62}{15}+2 a \coth(a) \right)
+ \frac{2}{\cosh^2(a)} \bigl(-1+a \tanh (a)\bigr) \right) \,, \nonumber \\
g_{aB}&=&  \frac{\pi}{\cosh^2 (a)} -
B\left(  \frac{a}{\sinh^2 (a)} \left( 3+\frac{7}{\sinh^2 (a)}
+\frac{5}{\sinh^4 (a)} \right) \right. \nonumber \\
&-& \left. \coth (a) \left(1+ \frac{11}{3\sinh^2 (a)}
+\frac{5}{\sinh^4 (a)}   \right) + \frac{a}{\cosh^2 (a)} + \tanh (a)
\right) \,, \nonumber \\
g_{BB}&=&\coth^2 (a) \left( \frac{4}{3}
+\frac{2}{\sinh^2 (a)}\bigl(-1+a\coth (a)\bigr)
+\frac{2}{\cosh^2 (a)}\bigl(-1+a\tanh (a)\bigr) \right) \,, 
\eea
\bea
\hspace*{-2.5cm}
V &=& \frac{8}{3 ( e^{4 a}-1)^3}
\left(24 a(1+3e^{4 a}) + 17 - 9 e^{4 a}-9 e^{8 a}+e^{12 a}\right)
+ 6 \pi B \bigl(\tanh (a)-1 \bigr)^2 \tanh (a)  \nonumber \\
&+& 2 B^2\frac{\coth ^2(a) }{(e^{4a}-1)^5}
\Bigl(12 a e^{2 a} (17-24 e^{2 a}+204 e^{4 a}-88 e^{6 a}
+246 e^{8 a}-72 e^{10 a}+44 e^{12 a}
\nonumber \\
&-& 8 e^{14 a}+e^{16 a}) - 9 + 208 e^{2 a}-155 e^{4 a}+736 e^{6 a}
-154 e^{8 a}-576e^{10 a} +234 e^{12 a}\nonumber \\
&-& 352 e^{14 a}+83 e^{16 a}-16 e^{18 a}+e^{20 a}\Bigr)
+ \frac{\pi}{16} B^3 \text{sech}^3(a)
\bigl(93 \sinh (a)+\sinh (3 a)-48 \cosh (a)\bigr) \nonumber \\ 
&+& \frac{4}{35} B^4 \frac{ \coth ^4(a)}{(e^{4 a}-1)^7}
\Bigl(840 a e^{4 a} (1-4 e^{2 a}+29 e^{4 a}-32 e^{6 a}+130 e^{8 a}
-56 e^{10 a}+130 e^{12 a} \nonumber \\
&-& 32 e^{14 a}+29 e^{16 a} - 4 e^{18 a}+e^{20 a})-1-28 e^{2 a}+987 e^{4 a}
-2688 e^{6 a}+14119 e^{8 a} - 7980 e^{10 a}
\nonumber \\
&+& 19915 e^{12 a}-19915 e^{16 a} + 7980 e^{18 a} -14119 e^{20 a}
+2688 e^{22 a}-987 e^{24 a}+28 e^{26 a}+e^{28 a}\Bigr) \, . 
\eea
Despite the denominator factors, these expressions are regular at $a=0$.
We
show the Ricci scalar curvature in Fig. \ref{fig:curv}. The curvature
for large $|a|$ matches that of the wormhole associated with a single
kink and is maximal at $A \approx -0.84$ ($B \approx \pm 0.84$).
Because there is a kink and antikink, the metric is
doubled and the curvature halved. The potential is shown 
in Fig. \ref{fig:pot}; it is asymmetric in $a$ and $B$.

\section{\label{sec:dynamics}Effective model dynamics}

\begin{figure}
\includegraphics[width=1.00\columnwidth]{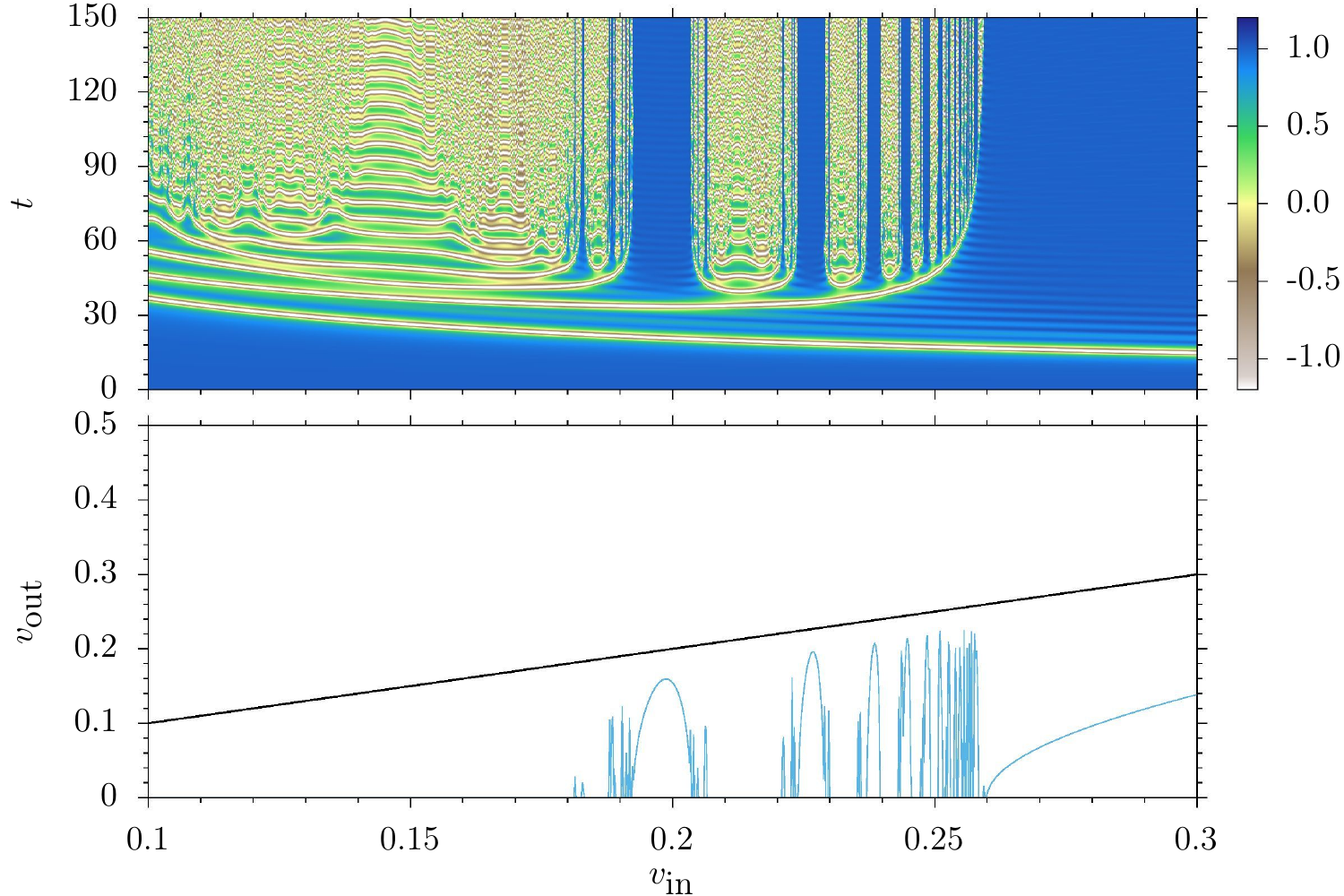}
\vspace*{-0.7cm}
\caption{KAK scattering in $\phi^4$ field theory}
\label{fig:full-dyn}
\vspace*{0.8cm}
\includegraphics[width=1.00\columnwidth]{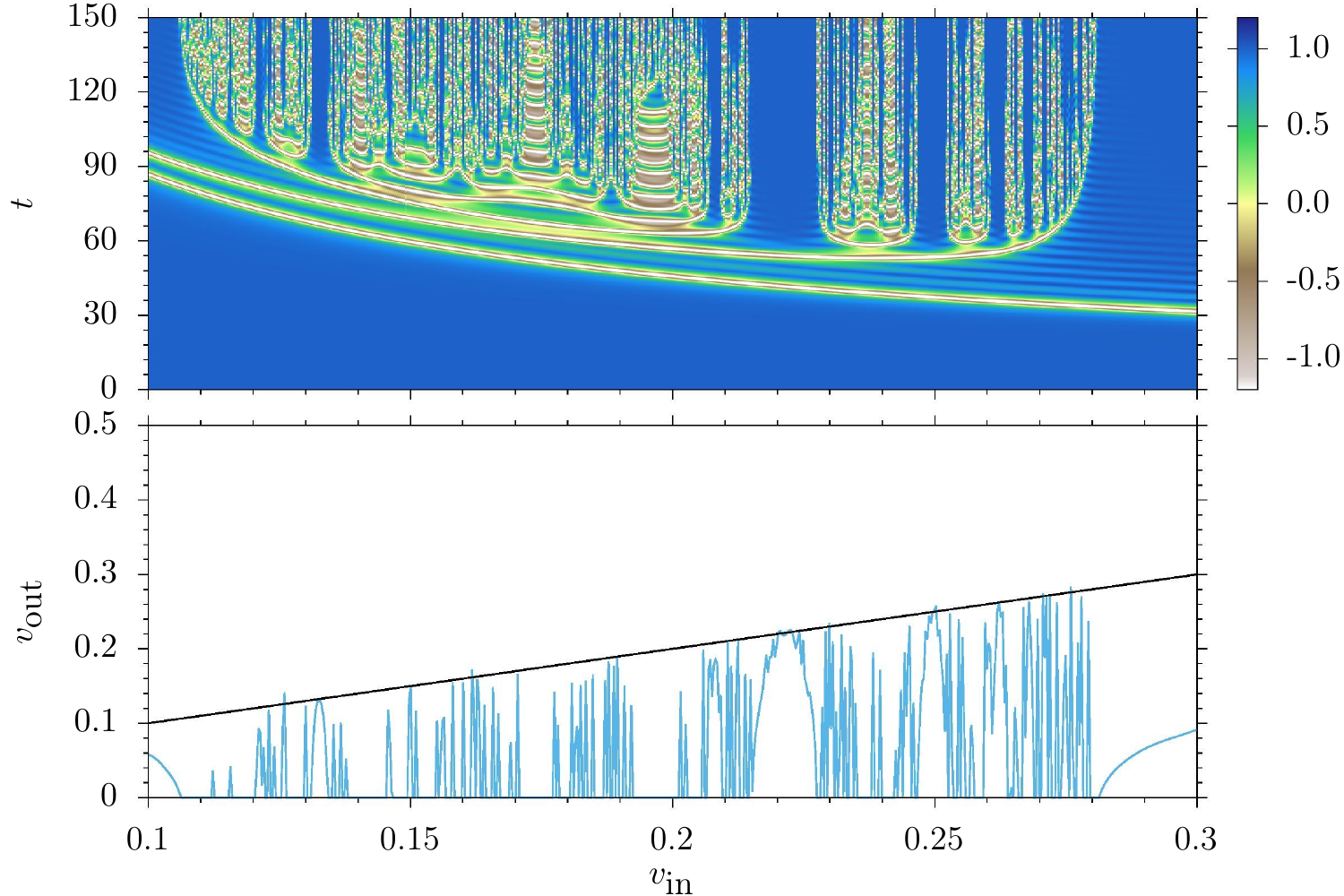}
\vspace*{-0.7cm}
\caption{KAK scattering in the effective model}
\label{fig:eff-dyn}
\end{figure}

Before discussing KAK collisions in the effective model, let us 
recall that in the field theory, the main feature is a fractal structure 
as a function of the initial velocity $v_{\rm in}$, distinguishing 
{\it annihilation to the vacuum} and {\it reflection} channels, see 
Fig. \ref{fig:full-dyn}. The figure shows the time evolution
of the field $\phi$ at $x=0$ and the final velocity $v_{\rm out}$ of the
outgoing kink, both as functions of the incoming kink velocity
$v_{\rm in}$. If the kink and antikink annihilate, $v_{\rm out}$ is
shown as zero. During annihilation, the incoming kink and antikink form a 
long-lived, quasi-periodic bound state, a {\it bion}, which slowly 
decays by emission of radiation. In the reflection channel, the kink 
and antikink perform a small number of {\it bounces}, then
reemerge and separate. The pattern of channels and of particular
$n$-bounce windows is fractal. For example, the first 2-bounce window
occurs for $ 0.1920 < v_{\rm in} < 0.2029$, and repeats infinitely
often as $v_{\rm in}$ increases. These windows are surrounded by 3-bounce 
windows, and this picture is replicated for higher-bounce windows. 
Bion formation occurs in the intermediate velocity intervals,
which appear in the figure as {\it bion chimneys}.
 
Overall, the fractal structure occurs in the approximate range
$v_{\rm in}=0.18$ -- 0.26. For smaller initial velocities there is always bion 
formation (a wide bion chimney) leading to annihilation, while for 
larger velocities there is just one bounce before the kink and 
antikink reflect back to infinity.

We now turn to the effective model (\ref{EffL}), defined on the moduli 
space with coordinates $(a,B)$. The dynamical ODEs require appropriate 
initial conditions. We assume the incoming kink and antikink approach 
symmetrically and are not vibrating, so impose relation 
(\ref{v-A}) between the initial shape mode amplitude and velocity. The
field configuration at each instant is given by (\ref{KAKrefined}).

In the effective model, there is similar field evolution and fractal
behaviour of velocities as in the field theory, including multi-bounce
windows, see Fig. \ref{fig:eff-dyn}. Fractal behaviour now occurs over a wider
range of velocities, up to $v_{\rm in} \simeq 0.282$. Because there is no
radiation mechanism, the outgoing velocities tend to be larger than in the
field theory. Despite this, the velocity patterns match
surprisingly well, apart from a small shift $\delta v_{\rm in} \simeq 0.02$.
The bion oscillations also match well.

It needs to be stressed that the fractal velocity structure has not
previously been reproduced in any effective model derived from the $\phi^4$
theory \cite{KG}. Phenomenological models revealing a chaotic
behaviour of the positions of the bounce windows and bion chimneys 
had little to do with the original theory and required arbitrary 
calibration of couplings.

\section{\label{sec:summary}Summary}

We have investigated an effective, collective coordinate model of symmetric
kink-antikink dynamics in $\phi^4$ theory -- a Lagrangian dynamics on a
curved 2-dimensional moduli space with a potential, where the coordinates
are the KAK separation and the (equal) amplitudes of the KAK shape
modes. Two crucial features are: {\it i)} initial conditions
including excitation of the kink shape modes, modelling Lorentz
contractions; {\it ii)} regularization of the moduli space metric
through use of an improved set of coordinates. There appears to be no
problem using a fixed (frozen) shape mode, even though such a mode can
disappear into the continuum spectrum as the KAK separation
decreases \cite{AORW}.

The model gives good results for the fractal velocity pattern of KAK
scattering, with its multi-bounce windows, and also for the field
evolution of bions, where the kink and antikink are captured. It would be
desirable to add a dissipation mechanism, modelling the coupling to
radiation, to have an upper bound on the number of bounces and for the
bion to decay to the vacuum.

\section*{Acknowledgements}

\vspace*{-0.4cm}

NSM has been partially supported by the U.K. Science and 
Technology Facilities Council, consolidated grant ST/P000681/1, and
thanks Maciej Dunajski for discussions about wormholes. 
KO, TR and AW were supported by the Polish National Science Centre, 
grant NCN 2019/35/B/ST2/00059.

\end{document}